\documentclass[a4 paper,11 pt]{article}
\pdfoutput=1
\usepackage{jheppub} 

\usepackage[T1]{fontenc}
\usepackage{amsmath}
\usepackage{physics}
\usepackage{cancel}
\usepackage{caption}
\usepackage{subcaption}
\usepackage{subcaption,longtable,stmaryrd}
\usepackage{comment}

\usepackage{subcaption}
\usepackage{float}
\usepackage{comment}
\DeclareSymbolFont{usualmathcal}{OMS}{cmsy}{m}{n}
\DeclareSymbolFontAlphabet{\mathcal}{usualmathcal}
\usepackage{bigints}
\usepackage{multicol}
\usepackage{blindtext}
\usepackage{tikz}
\usepackage{enumitem}
\usepackage{amsmath, amssymb, graphics, setspace}
\usepackage{float}
\newcommand{\mathsym}[1]{{}}

\newcommand{\unicode}[1]{{}}
\usepackage[mathscr]{euscript}
\DeclareSymbolFont{rsfs}{U}{rsfs}{m}{n}
\DeclareSymbolFontAlphabet{\mathscrsfs}{rsfs}
\DeclareSymbolFont{rmlargesymbols}{OMX}{mdbch}{m}{n}
\DeclareMathSymbol{\rmintop}{\mathop}{rmlargesymbols}{82}
\DeclareMathSymbol{\rmointop}{\mathop}{rmlargesymbols}{72}
\usepackage[customcolors,shade]{hf-tikz}

\usepackage{graphicx}

\definecolor{darkbrown}{rgb}{0.787, 0.26, 0.187}
\definecolor{alizarin}{rgb}{0.82, 0.1, 0.26}

\title{\textcolor{black}{Effects of monitoring on entanglement dynamics for $1+1$D $\mathbb Z_2$ lattice gauge theory
}}

\author[a]{Nilachal Chakrabarti,}
\author[b,c,d]{Nisa Ara,}
\author[a]{Neha Nirbhan,}
\author[a]{Arpan Bhattacharyya,}
\author[c,d]{and Indrakshi Raychowdhury}
\affiliation[a]{Department of Physics, Indian Institute of Technology Gandhinagar, Gujarat 382055, India.}

\affiliation[b]{Department of Theoretical Physics, Tata Institute of Fundamental Research,\\Homi Bhabha Road, Mumbai 400005, India}

\affiliation[c]{Department of Physics, Birla Institute of Technology and Science Pilani, Zuarinagar, Goa 403726, India.
}
\affiliation[d]{Center for Research in Quantum Information and Technology,
Birla Institute of Technology and Science Pilani, Zuarinagar, Goa 403726, India.}

\emailAdd{nilachalchakrabarti@iitgn.ac.in}
\emailAdd{nisa.ara@theory.tifr.res.in}
\emailAdd{neha.nirbhan1207@gmail.com}
\emailAdd{abhattacharyya@iitgn.ac.in}
\emailAdd{indrakshir@goa.bits-pilani.ac.in}

\abstract{The $(1+1)$-dimensional $\mathbb Z_2$ gauge theory is the simplest model that allows for quantum simulation to probe the fundamental aspects of a gauge theory coupled with dynamical fermions. To reliably benchmark such a system, it is crucial to understand the non-unitary quantum dynamics arising from effective non-Hermitian evolution  \textcolor{black}{and post-selected monitoring protocols}. \textcolor{black}{This work focuses on the post-selected non-Hermitian filtering dynamics of a $\mathbb Z_2$ gauge theory, where the non-Hermitian terms are associated with local and non-local gauge-invariant operators naturally present in the theory. We interpret the resulting dynamics as post-selected filtering, where different operator sectors are coupled to loss channels
with different rates. This gives a unified framework for both the local electric flux and particle-number terms and the non-local mesonic hopping term. Tensor network calculations are performed to probe the effect of the filtering for larger lattice sizes (up to 256-site systems). Using Matrix Product State calculations, the dynamics of entanglement entropy are studied as a function of the filtering rate and the coupling constant. We find that, under both local and non-local filtering, the late-time saturation value of the bipartite entanglement entropy remains independent of system size, providing no evidence of  a measurement-induced phase transition-like phenomenon in the post-selected dynamics across the range of filtering strengths, evolution times, and system sizes considered here.}}

\makeatletter

\begin{document}

\maketitle
\section{Introduction}\label{section-1}

The confluence of quantum information theory, condensed matter, and high-energy physics has opened new avenues for understanding the fundamental properties of quantum many-body systems. The heart of this intersection lies in the study of entanglement, which has evolved from a particular feature of quantum mechanics to a powerful lens for characterising complex quantum many-body systems, including their phases, dynamics, and computational complexity. A particularly fruitful yet challenging frontier is the application of these concepts to lattice gauge theories (LGTs), the primary non-perturbative framework for studying fundamental interactions, from quantum chromodynamics to emergent phenomena in condensed matter systems \cite{RevModPhys.51.659,PhysRevLett.109.175302,Zohar_2016,Martinez2016,Yang2020,Zohar:2021nyc,Zhou2022,PhysRevLett.131.050401,Desaules2024ergodicitybreaking,Mueller:2024mmk,Zhang2025_confinement,Osborne2025_U1Gauge,QuantumGeometry2025,Halimeh:2025vvp}.
The $(1+1)$-dimensional $\mathbb Z_2$ gauge theory stands out as the simplest and most foundational LGT, serving as an
essential theoretical laboratory for complex behaviours like confinement. When coupled with dynamical matter, such as staggered fermions, its physics becomes significantly richer 
 \cite{RevModPhys.51.659,Wegner1971,Fradkin:1978th,FradkinShenker1979,SenthilFisher2000,PhysRevB.84.235148,Zohar_2016,Kormos2017,PhysRevLett.118.266601,Smith2018,FRANK2020135484,PhysRevLett.124.120503,Zohar:2016wmo,PhysRevLett.125.256401,2019NatPh..15.1168S,Surace2020,Banuls2020,10.21468/SciPostPhys.10.6.148,Halimeh:2021lnv,Mildenberger:2022jqr,PhysRevLett.131.081901,Kebric:2024yaa,Chen:2024oao,r6sr-dv13}. The presence of fermions introduces the phenomenon of string breaking: the confining flux tube between two static charges can break due to the creation of fermion-antifermion pairs from the
vacuum. This physical process has a direct and profound signature in the entanglement structure of the system’s states \cite{Pichler2016,knaute2024entanglement,Fromm:2023npm}. Defining and computing entanglement entropy in LGTs may become conceptually involved due to the gauge constraints and presence of a non-trivial centre, which dictates that the physical Hilbert space is not a simple tensor product of local degrees of freedom. Seminal works have established a rigorous framework for defining a physically meaningful, gauge-invariant entanglement entropy, often by considering the algebra of local observables \cite{PhysRevD.89.085012}. For one-dimensional systems, tensor networks, and specifically matrix product states (MPS), have proven to be an exceptionally powerful numerical tool. They provide an efficient representation of the low-entanglement states typical of LGT ground states and allow for the accurate simulation of their real-time dynamics \cite{Pichler2016,Banuls:2018jag,Magnifico:2019kyj,Irmejs:2022gwv,Mathew:2025fim}, making them ideal for investigations at the interface of quantum information science and strongly correlated quantum many-body systems such as gauge theories for large system sizes. For $(1+1)$-dimensional $\mathbb Z_2$ gauge theory coupled to fermionic matter, the Hamiltonian can be mapped to a spin model. This removes gauge redundancy and provides a convenient framework for numerical simulations, further simplifying the computation of entanglement entropy and other observables using tensor network methods to study the dynamics of the systems. One of the main goals of this paper is to study the entanglement dynamics of $(1+1)$-dimensional $\mathbb Z_2$ gauge theory in detail using the tools of tensor networks. 

Parallel to these developments, the study of monitored quantum systems where unitary evolution is interspersed with projective measurements has uncovered a new class of non-equilibrium phenomena. A central discovery is the measurement-induced phase transition (MIPT)
\cite{skinner2019measurementinduced,li2018quantum,li2019measurementdriven,chan2019unitaryprojective,boorman2022diagnostics,Biella2021manybodyquantumzeno,szyniszewski2020universality,barratt2022transitions,zabalo2022infinite,barratt2022field,PhysRevResearch.2.043072,zabalo2022operator,iaconis2021multifractality,Sierant2022dissipativefloquet,bao2020theory,choi2020quantum,szyniszewski2019entanglement,block2022measurementinduced,jian2020measurementinduced,agrawal2021entanglement,gullans2020scalable,sharma2022measurementinduced,zabalo2020critical,vasseur2019entanglement,li2021conformal,turkeshi2020measurementinduced,sierant2022universal,gullans2020dynamical,PhysRevB.107.L220201} \footnote{This list is by no means exhaustive. Interested readers are referred to references and citations of these papers.}. This transition describes a competition: unitary dynamics tends to spread information
and generate entanglement, leading to a volume-law scaling of entanglement entropy, characteristic of thermalization. Conversely, frequent local measurements extract information from
the system, collapsing the wavefunction and hindering entanglement growth, resulting in a low-entanglement, area-law scaling phase. The transition between these two regimes is a novel dynamical phase transition that occurs not in the state itself, but in its entanglement properties, and is invisible to local order parameters. This paradigm has been extensively studied in random unitary circuits and is beginning to be explored in more structured systems with conservation laws, which serve as a stepping stone towards the rigid constraints of gauge symmetry \cite{PhysRevB.105.064306,PhysRevB.111.094315}. There are recent studies \cite{Li:2026phj,Maeno:2026ooq,Rhodes:2026atz} that have discussed algorithms for constructing gauge-invariant states in quantum simulators. 

Despite the progress in these two fields, the interplay between gauge constraints, matter-field dynamics, and quantum measurements remains largely uncharted territory. LGTs present a unique setting where measurements can be performed on physical degrees of freedom (gauge,
matter, or combined), and the inherent non-local gauge constraints might fundamentally alter the nature of the MIPT. This leads to several compelling open questions: How does the presence of a local
gauge symmetry affect the universal properties of the MIPT? The entanglement entropy of a gauge theory, though subtle, may offer a unique probe of dynamical processes such as string breaking. It is still unexplored how the entanglement structure changes due to the measurement of non-local (physical) extended observables. How do the entanglement dynamics and effect of measurement on the same differ in the strong coupling (confining) and weak coupling regimes of the theory, where the energy scales of gauge and matter fields are starkly different?

\textcolor{black}{In this article, we take a step towards addressing these questions. We investigate one aspect of this problem by studying post-selected non-Hermitian filtering dynamics in the $(1+1)$-dimensional $\mathbb{Z}_2$ lattice gauge theory coupled to staggered fermions using tensor network methods. We study a class of post-selected non-Hermitian filtering dynamics in which the anti-Hermitian terms are associated with gauge-invariant operators naturally appearing in the theory. These include the local electric flux and particle-antiparticle number operators, as well as the non-local hopping term involving fermionic creation or annihilation operators acting on two sites and the holonomy of the gauge fields acting on the link joining those two sites. A useful way to interpret this dynamics is as a no-click-type postselected evolution \cite{PhysRevLett.68.580,PhysRevLett.70.2273,Plenio:1997ep}, in
which different eigenspaces, or sectors, of the chosen gauge-invariant operator are coupled to external loss channels with different rates. For local operators, such as the electric flux and particle-antiparticle number, this interpretation is
closely related to the no-click monitored dynamics studied in
\cite{PhysRevB.103.224210, 10.21468/SciPostPhys.14.5.138}. However, since our analysis also includes the non-local mesonic hopping operator, which is Hermitian and gauge-invariant but not a positive projector, we adopt this broader filtering interpretation throughout the paper. Thus, the anti-Hermitian terms should not be understood in general as symmetric continuous measurements of the corresponding Hermitian observables themselves. Rather, they define a unified post-selected non-Hermitian filtering protocol associated with gauge-invariant operators.}

We meticulously map out the
entanglement dynamics as a function of the non-Hermitian filtering rate and investigate whether it shows a MIPT i.e the scaling (w.r.t to the (sub)-system size) of its late-time saturation value transitions from a volume-law to an area-law. We also find that the entanglement dynamics exhibit completely different behaviour in the presence of filtering compared to the case without filtering. Furthermore, we systematically investigate this in both the strong and weak coupling regimes. Our work bridges the gap between high-energy physics phenomenology and the rapidly developing field of monitored quantum systems, shedding light on the intricate nature of quantum information in gauge theories.
The paper is organized as follows. In Sec.~(\ref{section-1}), we give a brief overview of $(1+1)$-dimensional $\mathbb Z_2$ gauge theory and discuss the structure of the underlying Hilbert space. We also detail the computation of the entanglement entropy for this system under filtering and discuss the numerical methods that we use throughout the paper. In Sec.~(\ref{section-3}), we present the analysis of entanglement entropy without any filtering. In Sec.~(\ref{section-4}),  we turn our attention to the dynamics of the entanglement entropy under filtering dynamics associated with both local and non-local gauge-invariant
operators, and in Sec.~(\ref{section-5}), we investigate the late-time saturation value of the entanglement entropy for both strong and weak coupling regimes. Finally, in Sec.~(\ref{section-6}), we summarize our results and point out some future directions. Various consistency checks and benchmarks for our numerical methods are provided in the Appendix~(\ref{app-A})\,.

\section{Theoretical and computational frameworks}\label{section-1}
\subsection{Brief overview of 1+1D $\mathbb Z_2$ gauge theory}\label{section-1a}

We begin by briefly reviewing the necessary details regarding our model considered in this paper. We focus on a $\mathbb Z_2$ gauge theory coupled with dynamical staggered fermions in $(1+1)$-dimensions. The Hamiltonian for the system on the L-site lattice with open boundary conditions is given as \cite{Wegner1971, FradkinShenker1979, Davoudi2024scatteringwave,Mildenberger:2022jqr}, 
\begin{equation}\label{H1}
 H = x\sum_{i=0}^{L-2} \left(\psi^{\dag}_{i}\tau^{X}_{i,i+1}\psi_{i+1}+ \text{h.c.} \right)
+ \mu \sum_{i=0}^{L-1} (-1)^{i}\psi^{\dag}_{i}\psi_{i}+ \sum_{i=0}^{L-2}\tau^Z_{i,i+1}\,, 
\end{equation}
where $\psi^{\dag}_{i} (\psi_{i})$ are fermionic creation (annihilation) operators acting on $i^{th}$ site and $\tau^{\alpha}_{i,i+1} (\alpha=X, Z$ denotes the corresponding Pauli matrices) are the elements of $\mathbb Z_2$ representing the holonomies of gauge fields on each link between sites $i,i+1$. In (\ref{H1}), the dimensionless parameters $x, \mu$ are related to the fermion mass $m$, coupling $g$ and lattice spacing $a$ as: $$ x=\frac{1}{a^{2}g^{2}}, ~~\mu=2\frac{m}{g}\sqrt{x}\,.$$ For a fixed $\frac{m}{g}$ and lattice volume $V=La$, the continuum limit for the theory is obtained by taking $a\rightarrow 0$, $L\rightarrow\infty$ followed by $x\rightarrow\infty$. We will set $\frac{m}{g}=1$ for all of our computations. Hence, the coupling $x$ is the only tunable parameter in this case.

Being a gauge theory, the Hamiltonian is associated with the Gauss law operators  $G_i$ defined at each site $i$ as:
\begin{equation}
   G_i = \tau^Z_{i-1,i} \tau^Z_{i,i+1} e^{-i\pi \left( \psi_i^\dagger \psi_i - \frac{1 - (-1)^i}{2} \right)}\,.
\end{equation}
The Hamiltonian in (\ref{H1}) commutes with all the local Gauss law operators
\begin{equation}
 ~~[H,G_i]=0\,, \forall i
\end{equation}
and hence keeps the dynamics of the theory gauge-invariant.  The physical Hilbert space $\mathcal H_{phys}$ of the theory contains states $\{|\psi\rangle_{phys}\}$ which are gauge invariant, i.e 
\begin{equation} G_i\ket{\psi}=\ket{\psi}. \label{phys-state}\end{equation}

A Jordan-Wigner transformation can be performed for the fermions, and the resulting Hamiltonian is given as: 
\begin{equation}\label{H2}
   H = x \sum_{i=0}^{L-2} \left( \sigma_i^{+} \, \tau^X_{i,i+1} \, \sigma_{i+1}^{-} + \text{h.c.} \right)
+ \mu\sum_{i=0}^{L-1}(-1)^{i}\sigma^{Z}_{i}+\sum_{i=0}^{L-2}\tau^Z_{i,i+1}
\end{equation} where $\sigma^{+}= \frac{1}{2}(\sigma^{X}+i\sigma^{Y})\,,\sigma^{-}= \frac{1}{2}(\sigma^{X}-i\sigma^{Y})\,.$
In (\ref{H2}), there exist two independent species of spins, $\tau$ corresponding to the two-component $\mathbb Z_2$ gauge fields defined on the links and $\sigma$ corresponding to fermions obtained via a Jordan-Wigner transformation. Both are represented by Pauli matrices.

\begin{figure}[h]
    \centering
    \includegraphics[width=0.75\linewidth]{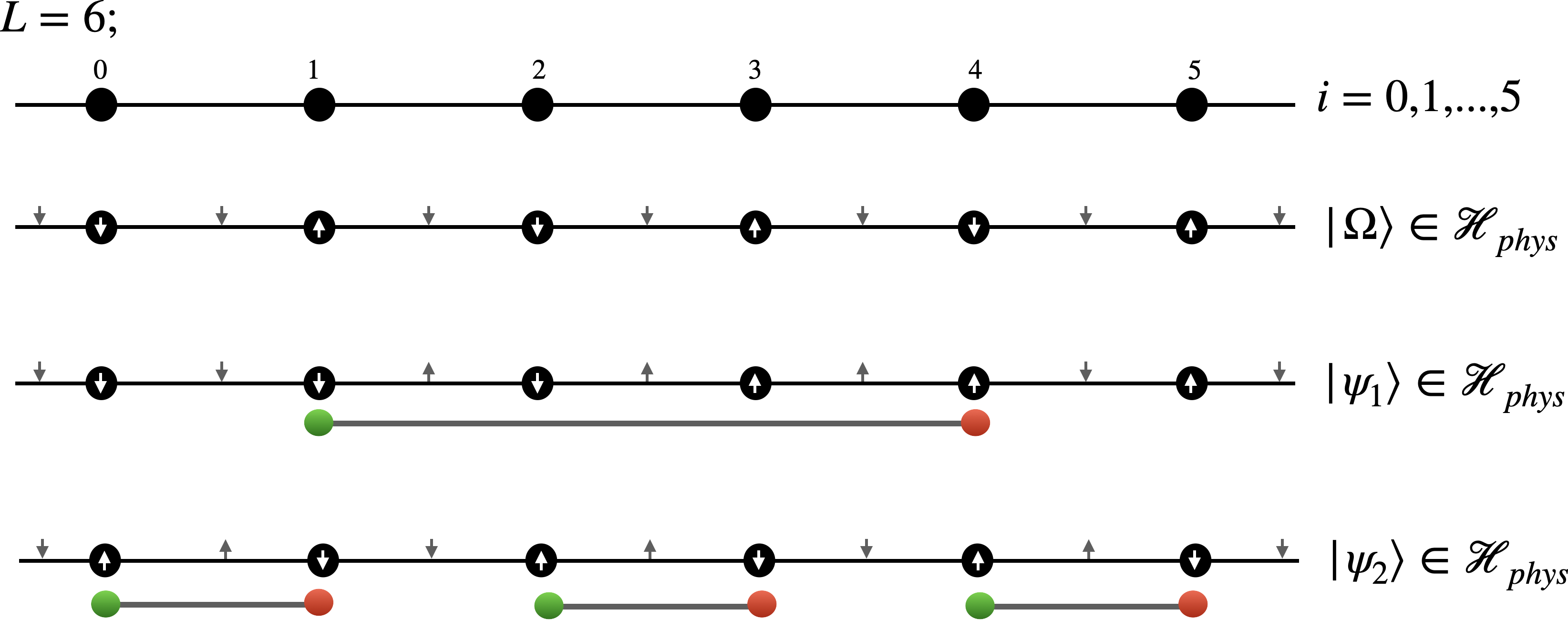}
    \caption{A diagram denoting physical states on a 6-site lattice. The topmost panel denotes the lattice and the label for each site;  the next one $|\Omega\rangle$ denotes the strong coupling vacuum - a global spin configuration on the lattice which corresponds to the presence of no particle, no anti-particle, and no gauge flux. The next two denote two global gauge invariant states $|\psi_1\rangle $ and $|\psi_2\rangle$, which contain 1 and 3 particle-antiparticle pairs, respectively, connected by gauge fluxes. Note that, for all three configurations, the incoming boundary fluxes are chosen as zero ($s=-1/2$), and for both the outgoing ones are also zero ($s=-1/2$) as the difference between the total number of particles and anti-particles contained in the lattice is zero. All three states correspond to a single global charge sector in the theory, with $s_{\textrm{global}}=-1/2$ \cite{Ara:2026o5}.} 
    \label{fig-1}
\end{figure}

The global matter-gauge Hilbert space is spanned by direct products of on-site spins and on-link spins over the entire lattice; however, the physical Hilbert space is a subset of that space in which all the states satisfy (\ref{phys-state}).
As dictated by the operator $G_i$, the on-site contribution to a physical state differs between even and odd sites where fermions and anti-fermions reside. We work with the convention that the presence (absence) of a fermion (anti-fermion) is mapped to the spin to be up $s=1/2$ and absence (presence) of the same is mapped to the spin to be down $s=-1/2$. Gauge invariance at each site implies that the spins on the links are flipped in the presence (absence) of a fermion (anti-fermion). In this work, we consider quench dynamics starting from a gauge-invariant state. Most commonly, we consider the \textit{strong coupling} vacuum to be a gauge-invariant initial state. For a staggered theory, the strong coupling vacuum is defined as a configuration where odd sites are filled with fermions and even sites are all empty, and no gauge flux is present across the lattice. In terms of Jordan-Wigner spins, it is obtained by assigning spin up to all odd sites and spin down to all even sites, while keeping all on-link gauge fields at $s=-1/2$. A pictorial description of the strong coupling vacuum and a couple of other gauge invariant states is given in Fig.~(\ref{fig-1}).  

Now that we have discussed the underlying Hilbert space structure and the initial state on which we will apply our quench protocol, we will outline the exact protocol, which includes \textcolor{black}{post-selected filtering process}. 

\subsection{A framework for monitoring Quantum Systems}

\textcolor{black}{As discussed earlier, measurements and monitoring play an important role in understanding quantum dynamics}. Recently, MIPT
\cite{skinner2019measurementinduced,li2018quantum,li2019measurementdriven,chan2019unitaryprojective,boorman2022diagnostics,Biella2021manybodyquantumzeno,szyniszewski2020universality,barratt2022transitions,zabalo2022infinite,barratt2022field,PhysRevResearch.2.043072,zabalo2022operator,iaconis2021multifractality,Sierant2022dissipativefloquet,bao2020theory,choi2020quantum,szyniszewski2019entanglement,block2022measurementinduced,jian2020measurementinduced,agrawal2021entanglement,gullans2020scalable,sharma2022measurementinduced,zabalo2020critical,vasseur2019entanglement,li2021conformal,turkeshi2020measurementinduced,sierant2022universal,gullans2020dynamical,PhysRevB.107.L220201} has been studied in the context of quantum spin-chain models and quantum circuits. These studies have demonstrated the existence of two phases (depending on the measurement rate): an error-correcting phase, in which the state remains resilient to external measurements, and a Zeno phase, in which the state dynamics collapse due to measurements. \textcolor{black}{Motivated by these developments, we study how analogous non-unitary, post-selected dynamics affect the entanglement structure of the $\mathbb Z_2$ lattice gauge theory described in Sec.~(\ref{section-1a})}.

When a quantum system is continuously monitored by a measurement operator, its evolution is not deterministic and is described by the Stochastic Schrodinger equation \cite{carmichael2009open, wiseman2009quantum}, which includes the effect of measurement on the system. When a quantum state is being measured, a competition used to happen between unitary evolution and measurement. The evolution can occur in two ways: first, through a quantum jump; and second, slowly and continuously under measurement. The second scenario is called the \textit{no-click} limit  \cite{wiseman2009quantum,carmichael2009open, gardiner2004quantum,daley2014quantum,Jacobs2014,PhysRevB.105.205125}. In the `no-click' limit, starting from an initial state, one evolves the system using an \textit{effective non-Hermitian Hamiltonian}, \textcolor{black}{where the
non-Hermitian terms are associated with measurement operators} and the measurement rate determines the non-Hermitian strength. \textcolor{black}{Measurement-induced transitions have also been studied in such post-selected non-Hermitian settings \cite{10.21468/SciPostPhys.14.5.138, 10.21468/SciPostPhysCore.6.3.051, PhysRevB.103.224210, Chakrabarti:2025hsb} for certain spin-chain models. In this work, we adopt a unified post-selected non-Hermitian filtering framework
for the $\mathbb Z_2$ gauge theory. The anti-Hermitian terms are associated with gauge-invariant operators of the theory, and the parameter $\gamma$ controls the corresponding filtering strength. Equivalently, the dynamics may be viewed as a no-click-type evolution in which different eigenspaces or sectors of the chosen
gauge-invariant operator are coupled to external loss channels with different rates. For the local electric-flux and particle-antiparticle number operators, this connects closely with the usual no-click interpretation used in monitored spin or fermionic
chains. However, since we also consider the non-local mesonic hopping operator, whose direct interpretation as a symmetric continuous measurement is more subtle, we use the
broader terminology of post-selected filtering throughout. This deterministic non-Hermitian evolution allows us to study the entanglement dynamics using tensor network methods easily, while leaving the analysis of the full stochastic monitored dynamics
for future work.}

\textcolor{black}{To study the above-mentioned post-selected non-Hermitian dynamics, we have added three different terms- 
(to be explicitly defined in  the subsequent sections): \textit{electric field}, \textit{staggered mass term}, and \textit{hopping} with 
this simple model of $\mathbb{Z}_{2}$
gauge theory coupled with dynamical matter}. We start from a strongly coupled vacuum state as defined in Sec.~(\ref{section-1a}) and construct the density matrix $\rho(0)$. Then we time evolve it in the following way,  \begin{equation}\label{denmat eqn}
    \rho(t) = \frac{e^{-iH_{\textrm{eff}}\,t}\rho(0)e^{iH^{\dagger}_{\textrm{eff}}\,t}}{\textrm{Tr}\left(\textcolor{black}{e^{-iH_{\textrm{eff}}\,t}\rho(0)e^{iH^{\dagger}_{\textrm{eff}}\,t}}\right)}
\end{equation}
where $H_{\textrm{eff}}$ is of the form, $H_{\textrm{eff}}= H_{0}-i\gamma H_1$, where $H_{0}$ is defined in (\ref{H2}) and \textcolor{black}{$H_1$ denotes the part that gives rise to the non-hermitian dynamics}. 
$\gamma$ denotes the non-Hermitian filtering rate. The denominator in (\ref{denmat eqn}) makes sure that the state remains normalized at each point in time. Due to the simplification introduced, we can study the dynamics using the computational framework of matrix product state (MPS) described in the next subsection.

\begin{figure}[htb!]
\centering
\includegraphics[width=0.8\linewidth]{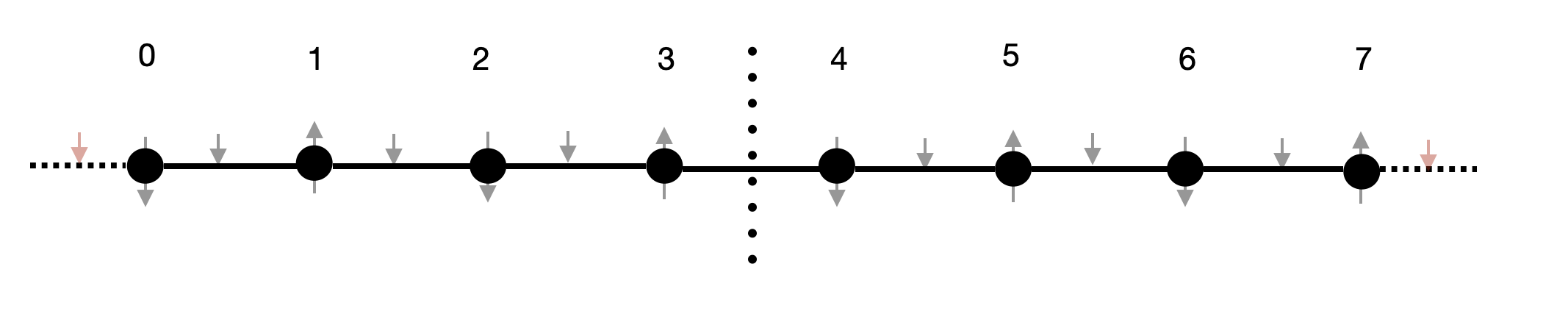}
\caption{\textcolor{black}{A representative diagram of a physical state on an 8-site lattice. The dotted line marks the bipartition used to compute the entanglement entropy, with subsystem $A$ on the left and its complement $B$ on the right. Local filtering terms, associated with site or link operators, are applied throughout the system. For the non-local mesonic hopping operator, which acts on two neighbouring matter sites and the intervening gauge link, we mainly restrict the filtering terms to one subsystem so that they do not cross the entanglement cut} \cite{Ara:2026o5}.}
\label{fig for bipartition}
\end{figure}

In this paper, we will study the dynamics of the entanglement. For that, we will compute the entanglement entropy under this non-Hermitian evolution. We start by dividing the system into two subsystems. Then, we calculate the reduced density matrix $\rho_A(t)$ of subsystem A by tracing out subsystem B. Fig.~(\ref{fig for bipartition}) demonstrates this bi-partitioning for our case, including how the non-Hermitian part of the effective Hamiltonian are acting on the (sub)-systems.
Using $\rho_A(t)$, we compute the von Neumann entropy for subsystem A and denote it  $S(L/2,t)$. \begin{equation}S(L/2,t) =-\textrm{Tr}\Big(\rho_{A}(t)\ln \rho_A(t)\Big)\,. \label{ent}
\end{equation} 

As mentioned earlier, the Hamiltonian for the $(1+1)$-dimensional $\mathbb Z_2$ gauge theory coupled to fermionic matter can be mapped to a spin model, which further simplifies the computation of entanglement entropy, as we can use (\ref{ent}). To investigate the presence of possible MIPT, we need to determine the scaling of $S(L/2,t)$ with system size, and it is crucial to probe systems of very large size. We achieve this by using tensor network tools as described below.

\subsection{Computational Framework}

We now briefly describe the tensor network methods used for entanglement entropy and study its dynamics in this paper. As we focus on $(1+1)$-dimensions, the matrix product state (MPS) is a well-established tool that efficiently represents the quantum states of the system, especially when the system is large and exact diagonalization becomes exponentially expensive \cite{Orus:2013kga, Bridgeman:2016dhh}. We use the ITensor package \cite{itensor} for all our numerical analysis. Matrix Product States (MPS) provide a remarkably efficient way to simulate the dynamics of large, strongly-coupled one-dimensional quantum systems. Instead of storing the full quantum state vector, whose size grows exponentially with the number of particles, MPS represents the state as a chain of interconnected smaller tensors. The amount of entanglement is controlled by a parameter called the bond dimension. As long as the entanglement in the system remains limited (as reflected in the analysis presented in the later part of the paper), the required bond dimension stays small. This keeps the computational cost of simulating time evolution manageable, scaling polynomially rather than exponentially with system size, and enabling accurate calculation of the dynamics of hundreds of quantum particles. 

MPS calculations usually start by calculating the ground state of the Hamiltonian, where the Hamiltonian is written in Matrix Product Operator (MPO) form \cite{Orus:2013kga,itensor}, and the ground state is obtained via DMRG (Density Matrix Renormalization Group) technique, starting from an initial guess wavefunction written in MPS form. A desired initial state can also be mapped to its MPS representation to study its real-time evolution.
After evolving the quantum state to a desired time using the Matrix Product Operator (MPO) representation of the $\mathbb{Z}_2$ gauge theory Hamiltonian, the resulting state is naturally encoded as an entangled Matrix Product State (MPS). \textcolor{black}{All real-time evolutions reported in this paper were performed using second-order time-evolving block decimation (TEBD) \cite{itensor}, implemented in ITensor with a second-order Suzuki–Trotter decomposition.} Finally, we compute the entanglement entropy. Extracting it from this MPS is incredibly efficient. To do this, the system is first divided into two subsystems, A and B, corresponding to cutting a single `bond' from the chain of tensors that forms the MPS. The magic of the MPS formalism is that the singular values ($\lambda_i$) across this bond are precisely the Schmidt coefficients of the quantum state's bipartition. Therefore, instead of needing to construct the full, exponentially large reduced density matrix, one needs to perform a singular value decomposition (SVD) at the bond of interest. The von Neumann entanglement entropy defined in (\ref{ent}) is then calculated directly from these singular values using the formula \textcolor{black}{$S(L/2,t) = -\sum_i \lambda_i^2 \ln{(\lambda_i^2)}$.} This procedure is computationally inexpensive and a native operation in tensor network algorithms, making it a powerful tool for tracking entanglement dynamics in real time.

Now we are ready to discuss the dynamics of the entanglement entropy for $(1+1)$-dimensional $\mathbb{Z}_2$ gauge. We first discuss the case with no filtering, and then contrast it with the case with non-Hermitian filtering. 

\section{Dynamics of Entanglement entropy of $\mathbb Z_2$ gauge theory without any filtering}\label{section-3}

In this section, we investigate the dynamical structure of entanglement entropy (EE) for a $\mathbb{Z}_2$ gauge theory coupled with dynamical staggered fermions in $(1+1)$-dimensions without any filtering terms. As discussed earlier, we pick a strongly coupled vacuum (Fig.~(\ref{fig-1})) as our initial state $|\psi(t=0)\rangle\,.$ Here, all the matter (antimatter) sites are occupied by down (up) spin, and the links are occupied by down spin. The unitary time evolution of this state is as follows $\ket{\psi(t)}= e^{-iHt}\ket{\psi(0)}$
where $H$ is the Hamiltonian defined in $(\ref{H2})$. We bipartition the system by making a partition at the $L/2$ site (cutting the system through the link joining the sites $L/2$ and $L/2+1$ as shown in Fig.~(\ref{fig for bipartition}). 
The von Neumann entropy of the subsystem is then given by (\ref{ent}).

As mentioned earlier, we have performed the real-time evolution using tensor network methods in the Matrix Product State (MPS) representation, as implemented in ITensors \cite{itensor}. To approximate the time-evolution operator, $e^{-iH\,\delta}$, we primarily use a second-order Trotter decomposition acting on non-overlapping three-site clusters. This procedure introduces a local truncation error of order $\delta^3$ (global truncation error of the order of $\delta^{2}$) per step. We carry out simulations on a chain of $L$ sites with a time step of $\delta = 0.1$ and a total evolution time of $T = 100$. We allow the MPS bond dimension to grow up to $\mathcal{D} = 1000$ with a truncation cutoff of $\chi = 10^{-8}$. We choose the strongly coupled vacuum state as the initial state and evolve it in time, normalizing the state at each time step. For all subsequent calculations, we will use these parameters. Also, we have explicitly checked that Gauss's Law holds throughout the duration of our observations. This is an essential point, as we do not want to spoil the underlying Gauge structure while we are investigating the dynamics.

\subsection*{Entanglement dynamics:} 

\begin{figure}[H]
    \centering
    \subcaptionbox{The plot depicts the evolution of EE without any non-Hermitian filtering. EE shows no saturation even at late times.}[0.46\linewidth]{\includegraphics[width=1.1\linewidth]{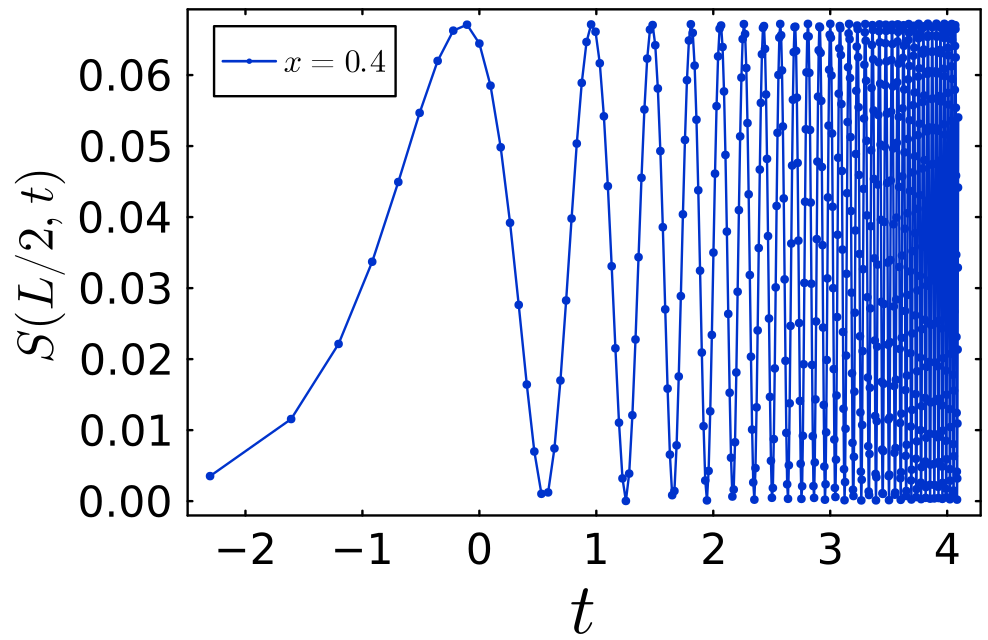}}
    \hspace{0.05\linewidth}
    \subcaptionbox{Time averaged entanglement entropy for different values of $x$ .}[0.46\linewidth]{\includegraphics[width=1.1\linewidth]{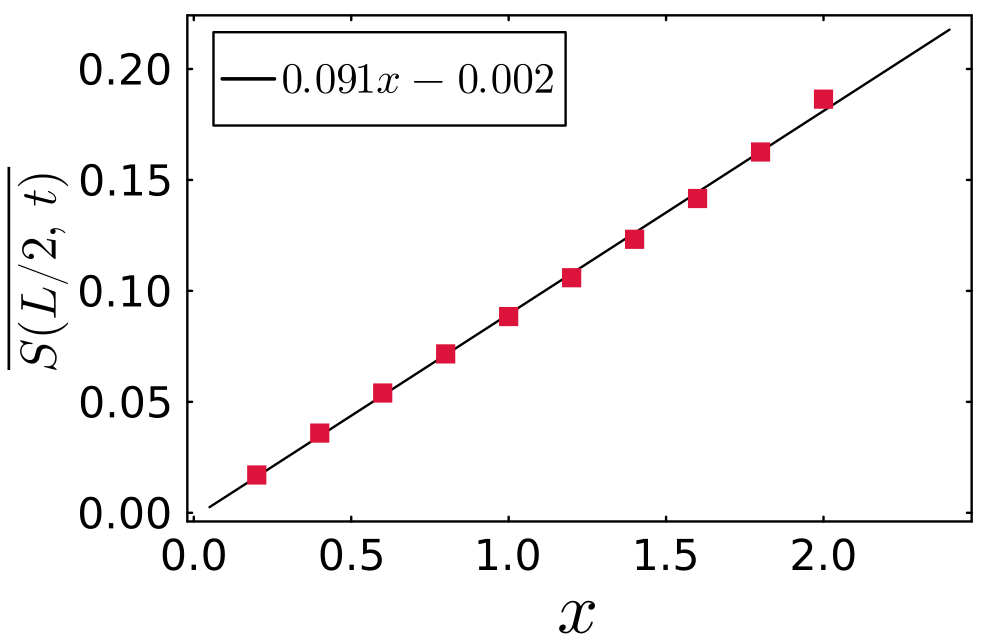}}
    \caption{Time evolution of the entanglement entropy for a system size of $L=32$ \cite{Ara:2026o5}.}
    \label{figent}
\end{figure}

The dynamical behaviour of entanglement entropy is shown in Fig.~(\ref{figent}) ({\bf left panel}). We observe that EE doesn't saturate for our theory. EE increases with time and then oscillates at late times. This behaviour persists for different values of the coupling parameter $x\,.$ It is interesting to note that the absence of saturation (thereby possibly thermalization \cite{Yang:2025vjh}) of EE  for $\mathbb{Z}_2$ theory in $(1+1)$-dimensions in the absence of any non-Hermitian filtering terms if one starts from a strongly coupled vacuum state. Furthermore, we plot the time-average EE as a function of the coupling parameter $x$, and it continues to grow as $x$ increases (towards the continuum limit). This is also shown in Fig.~(\ref{figent}) ({\bf right panel}).

\section{\textcolor{black}{Post selected dynamics 
in $\mathbb{Z}_2$ gauge theory: Effect on entanglement}}
\label{section-4}

\textcolor{black}{We now turn to the effect of post-selected non-Hermitian filtering on the dynamics of entanglement entropy in the $\mathbb{Z}_2$ gauge theory. Our motivation is twofold: (a) to
understand whether the entanglement entropy saturates under such filtering dynamics, and (b) whether we observe a MIPT-like transition.
We consider filtering dynamics associated with three gauge-invariant operators: the electric flux, the particle-antiparticle number, and the mesonic hopping operator. The first two are local operators, while the last one is a non-local extended operator involving two matter sites and the gauge link joining them. Although the local cases are closely connected to the no-click limit of the monitored system, we use a unified interpretation for all three cases as post-selected filtering with respect to gauge-invariant operator sectors. This is particularly useful because the hopping operator is not of the form of a projector and therefore does not admit an immediate
interpretation as a standard continuous measurement channel. Most previous studies of MIPT have focused on local measurements in spin chains or quantum circuits. In contrast, the present gauge-theory setting naturally contains both local gauge-invariant operators and extended non-local operators. This allows us to study, within a unified post-selected filtering framework, how the locality structure of the operator entering the non-Hermitian term affects the entanglement dynamics. Hence, this makes the current study all the more exciting. In the next section, we investigate the post-selection of local operators}.

\subsection{\textcolor{black}{Post selected dynamics with respect to local observables: electric flux and particle-antiparticle number}}
\label{section-4a}

\subsection*{Electric flux:}
Starting from the strongly coupled vacuum state, \textcolor{black}{we will study the post-selected filtering dynamics for electric flux operators $\tau^Z_{j,j+1}$ acting on the link between sites $ j$ and $ j+1$}. As we will work in the no-click limit, the  effective Hamiltonian, which will govern the time-evolution, is given by,
\begin{equation}
\label{H3}
   H_{\textrm{eff}} = x \sum_{j=0}^{L-2} \left( \sigma_j^{+} \, \tau^X_{j,j+1} \, \sigma_{j+1}^{-} + \text{h.c.} \right)
+ \mu\sum_{j=0}^{L-1}(-1)^{j}\sigma^{Z}_{j}+\sum_{j=0}^{L-2}(1-i\gamma)\tau^Z_{j,j+1}
\end{equation}
where $\gamma$ is the filtering strength. We evolve the initial density matrix $\rho(0)$ according to the equation (\ref{denmat eqn}) and compute the EE. 
\subsection*{Particle-antiparticle number:}
\textcolor{black}{We also study the post-selected filtering dynamics for another local operator i.e the particle-antiparticle number ($(-1)^{j}\sigma^{Z}_{j}$)}.
The non-Hermitian Hamiltonian in this case is given by,
\begin{align}
\begin{split}
 H_{\textrm{eff}} = x\sum_{j=0}^{L-2} \left( \sigma_j^{+} \, \tau^X_{j,j+1} \, \sigma_{j+1}^{-} + \text{h.c.} \right)
+(\mu -i\gamma)\sum_{j=0}^{L-1}(-1)^{j}\sigma^{Z}_{j}+\sum_{j=0}^{L-2}\tau^Z_{i,i+1}\,.
\end{split}
\end{align}
Again, $\gamma$ denotes the filtering strength. 
 
 Before proceeding to discuss our results, let us briefly discuss how we have benchmarked our tensor-network codes to showcase the validity of our results. 

\subsection*{Benchmarking the code:} 

We benchmark our tensor network code against an exact-diagonalization code for a small system size. Furthermore, we have also performed the time evolution (for that small system size) using ITensor by implementing three methods: the \textcolor{black}{1st- and 2nd-order TEBD, as well as the time-dependent variational principle (TDVP) \cite{PhysRevB.102.094315}}. The $S(L/2,t)$ obtained from all these methods matches with each other over the entire time range. This is demonstrated in Fig.~(\ref{Fig-App1}) of Appendix~(\ref{app-A}). Furthermore, we have checked the behaviour of $S(L/2,t)$ by changing the bond dimension $\mathcal{D}$ and the truncation cutoff $\chi\,.$ Again, we find perfect agreement. This is demonstrated in Fig.~(\ref{cut-off}) and Fig.~(\ref{EE_subs}) of Appendix~(\ref{app-A}). These help us perform consistency checks (and test the convergence of our results) by ensuring that our numerical results are independent of the choice of various intrinsic ITensor parameters and depend only on the parameters of the underlying theory. 

\begin{figure}[htb!]
    \centering
  \subcaptionbox{Electric flux\,.}[0.46\linewidth]{\includegraphics[width=1.1\linewidth]{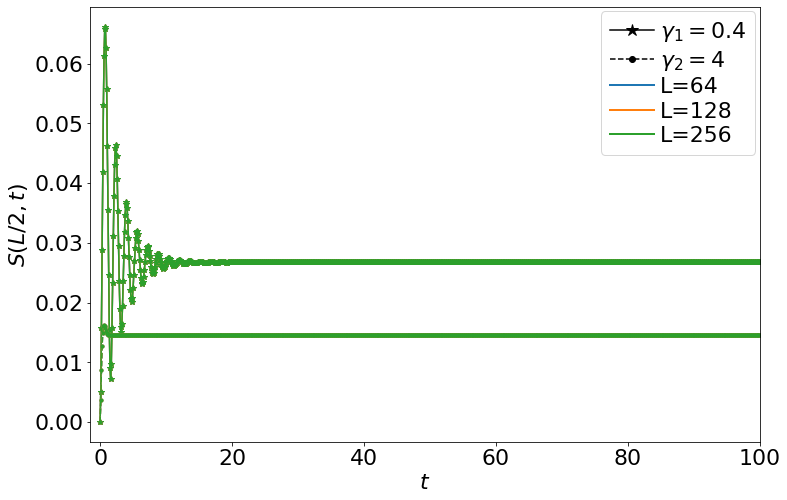}}
    \hspace{0.05\linewidth}
    \subcaptionbox{Particle-antiparticle density\,.}[0.46\linewidth]{\includegraphics[width=1.1\linewidth]{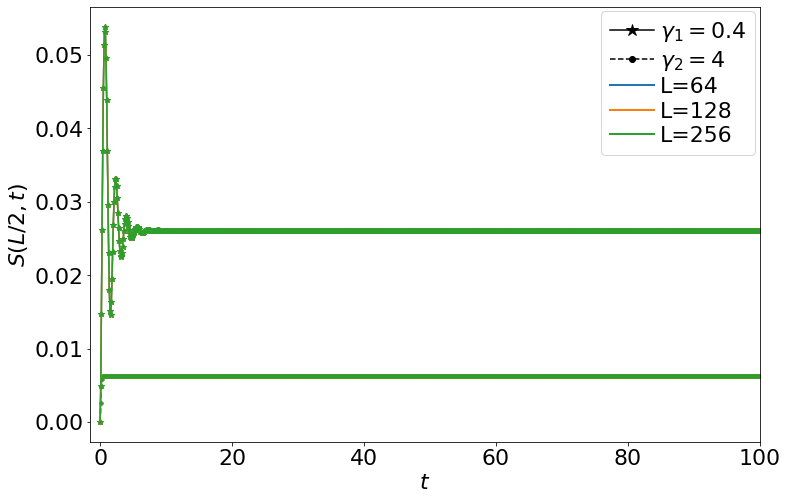}}
    \caption{Entanglement dynamics \textcolor{black}{while post-selecting with respect to} the local observables for two different values of $\gamma$ with $x=0.5\,.$}
    \label{Fig-4}
\end{figure}

\subsection*{Entanglement dynamics:}

Now we will present our results. In Fig.~(\ref{Fig-4}), we plot the evolution of EE over time for various system sizes and filtering strengths. Several features emerge from these plots; we list them below.
\begin{enumerate}
    \item \textcolor{black}{The plots clearly indicate that, regardless of the system size, the qualitative behavior of the evolution of EE, for the two local filtering terms considered here i.e electric field flux and density of particle-antiparticle pairs, is the same. It perhaps indicates that the dynamical evolution of EE depends mainly on whether the operators entering the filtering terms are local or non-local. In the next subsection, we contrast this local-operator behaviour with the dynamics obtained from filtering associated with the non-local mesonic hopping operator}.
    \item Unlike the case when there is no filtering (i.e $\gamma=0$), the EE \textit{saturates at late times} after showing oscillations at early times. Moreover, it reaches a lower saturation value with increased filtering strength, as depicted in Fig.~(\ref{Fig-4}). We will shortly conduct a thorough investigation into the dependence of the late-time saturation value of EE on the filtering rate $\gamma\,.$ 
    \item Interestingly, we find that the late-time saturation value of EE is independent of system size (here, we consider three lattice sizes: $L=64, 128, 256$). Although we show it for two particular values of filtering strength in the figure, we have scanned extensively over the values of $\gamma$ for mixed values of $x$ and vice versa. This remains true. Hence, we find no evidence of a MIPT-like finite-size scaling transition within the range of filtering strengths, evolution times, and system sizes investigated here. 
\end{enumerate}   
We also note that we have repeated this exercise with a different initial state to check whether the above-mentioned results are sensitive to the choice of initial state. In the Appendix~(\ref{app-B}), we have repeated this same exercise for a different initial state, namely the ground state of the $\mathbb{Z}_2$ Hamiltonian as defined in (\ref{H2}). We find the same behaviour as reported above.

\subsection*{Dependence of late-time saturation value of EE on filtering rate:}
From the previous Fig.~(\ref{Fig-4}), we found that entanglement entropy exhibits no scaling behaviour (w.r.t to the system size) for different local filtering terms. We now set the lattice size to $L=64$ and $x=0.5\,.$ \textcolor{black}{ We now study the dependence of EE on the filtering strength $\gamma\,.$}

\begin{figure}[H]
 \centering
 \subcaptionbox{Electric flux filtering\,.}[0.49\linewidth]{\includegraphics[width=\linewidth]{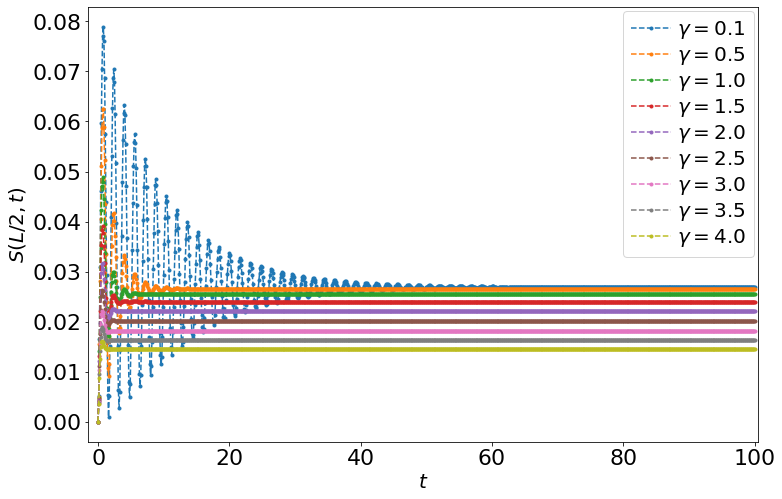}} 
 \label{Fig6a}
 \hfill
 \subcaptionbox{The functional form of the saturation value of EE with respect to filtering strength $\gamma\,,$ $f(\gamma)=ae^{-b\gamma}+c \gamma + d\,, $ with {$a=-0.00329217,b=1.64527, c= -0.00391089,d = 0.0301029$} evaluated c $t_{\textrm{sat}}=100$ for all values of $\gamma$ considered here.}[0.48\linewidth]{\includegraphics[width=\linewidth]{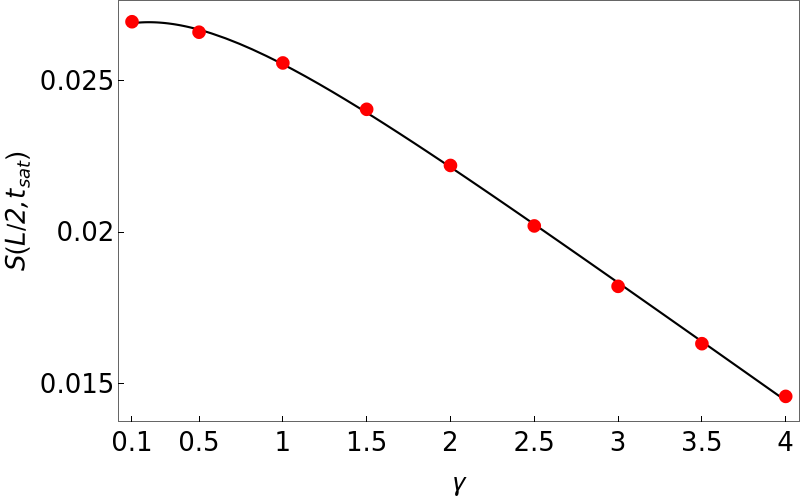}} 
 \label{Fig6b}
 \caption{Entanglement dynamics \textcolor{black}{while post-selecting with respect to the electric field operator} for different filtering rates for $x=0.5$ and $L=64\,.$}
 \hfill
\label{Fig-6}
\end{figure}

\begin{figure}[H]
 \centering
 \subcaptionbox{Particle-anti particle number filtering\,.}[0.49\linewidth]{\includegraphics[width=1.0\linewidth]{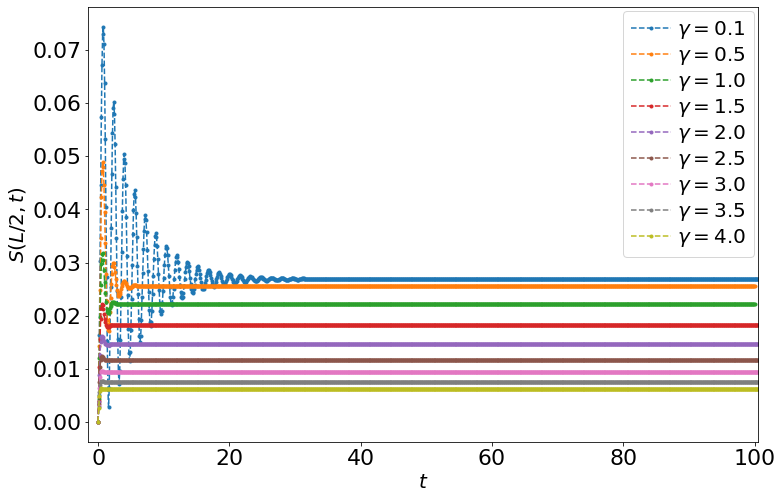}} \label{fig7a}
 \hfill
 \subcaptionbox{The functional form of the saturation value of EE with respect to filtering strength $\gamma\,,$ $f(\gamma)=ae^{-b\gamma}+c\,,$ with 
 {$a= 0.0437349,b=0.188269,c=-0.0149996$} evaluated at a late time $t_{\textrm{sat}}=100$ for all values of $\gamma$ considered here.}[0.48\linewidth]{\includegraphics[width=1.0\linewidth]{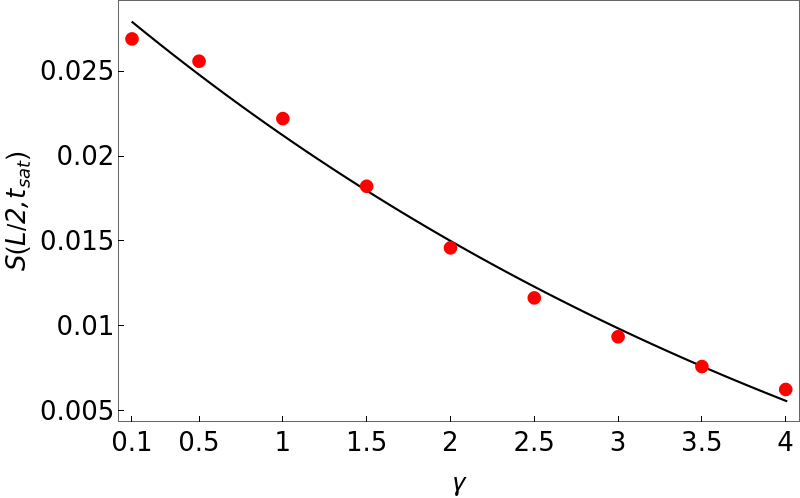}} \label{fig7b}
 \hfill

\caption{Entanglement dynamics \textcolor{black}{while post-selecting with respect to the particle-antiparticle number} for different filtering rates for $x=0.5$ and $L=64$.}
\label{Fig-7}
\end{figure}

From Figs.~(\ref{Fig-6}) and (\ref{Fig-7}), we observe that the EE saturates after some initial oscillations on a longer timescale. With increasing $\gamma$, the oscillations decrease and the late-time saturation value of the EE $S(L/2, t_{\textrm{sat}})$ decreases, showing  Quantum Zeno-like effect \cite{PhysRevLett.86.2699,skinner2019measurementinduced, Biella2021manybodyquantumzeno, li2018quantum, PhysRevB.111.094315}. Moreover, we find the functional form of the late-time saturation value of EE as a function of filtering strength $\gamma\,.$ This is shown in the {\bf right panels} of Figs.~(\ref{Fig-6}) and (\ref{Fig-7}). Note that we have done this analysis for $x <1\,.$ Keeping in mind the continuum limit (as discussed in Sec.~(\ref{section-1})), we repeat this same exercise for $x> 1$ in Appendix~(\ref{c}). We find the same Zeno-like behaviour, although the functional form of $S(L/2, t_{\textrm{sat}})$ is different compared to the $x<1$ case, as is evident from the plots in the right panels of Figs.~(\ref{Fig-AppD1}) and (\ref{Fig-AppD2}).

\subsection*{Entanglement entropy under local post-selected filtering in the strong and weak coupling regime}\label{section-5}
We have shown that the late-time saturation value of EE remains independent of system size for both of the local filtering terms considered previously for various values of the filtering rate $\gamma\,.$ Finally, we study the dependence of EE on the coupling parameter for fixed values of $\gamma\,.$ 
 We fix the system size to be $L=64\,.$ and compute $\textrm{EE}$ for different values of coupling parameter $x$ \textcolor{black}{while post-selecting with respect to the local observables}. We show the results for two different filtering rates.
\begin{figure}[H]
    \centering    \subcaptionbox{$\gamma=0.5$}[0.47\linewidth]{\includegraphics[width=1.1\linewidth]{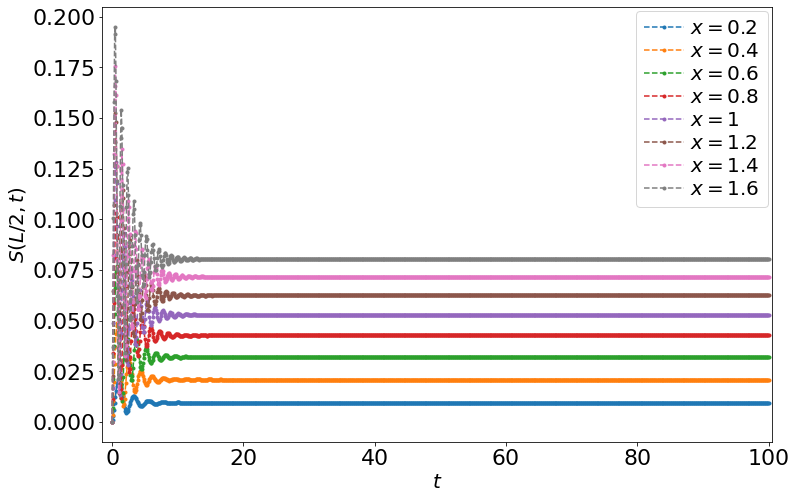}} 
    \label{fig9a}
    \hfill
    \subcaptionbox{$\gamma=1.5$}[0.47\linewidth]{\includegraphics[width=1.09\linewidth]{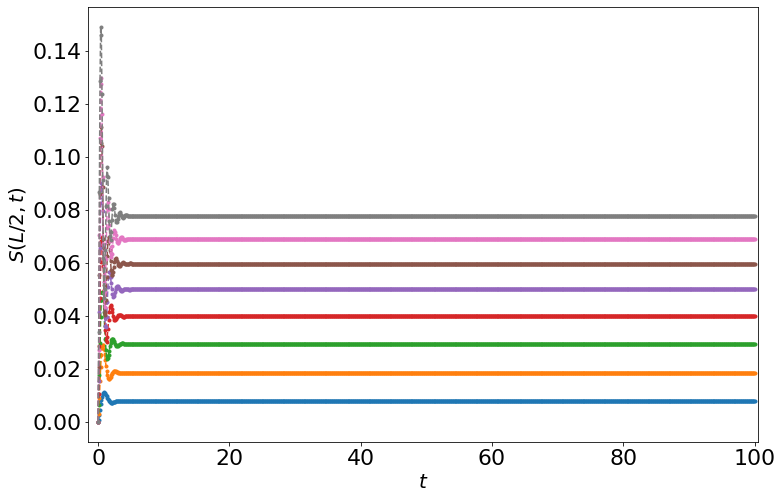}}
    
    \label{fig9b}
    \hfill
   \caption{Saturation value of EE versus $x$ under electric-flux filtering.} 
    \label{Fig-9}
\end{figure}

\begin{figure}[H]
   \centering    \subcaptionbox{$\gamma=0.5$}[0.47\linewidth]{\includegraphics[width=1.1\linewidth]{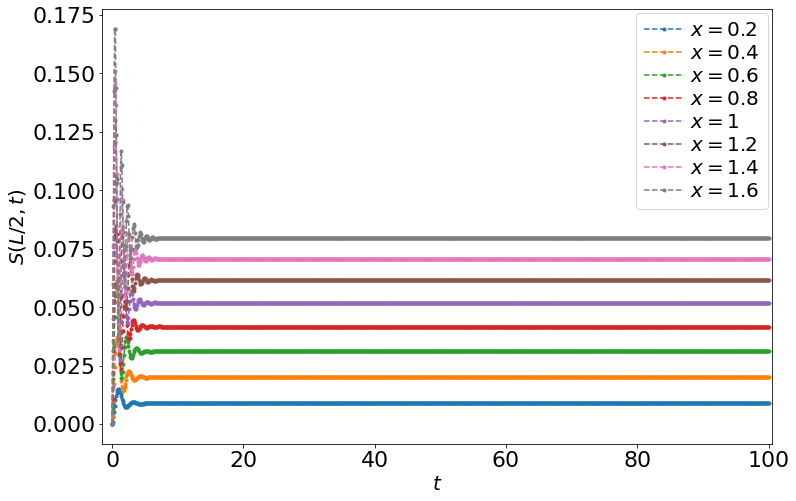}} 
   \label{fig10a}\hspace{0.10 cm}
   \hfill
    \subcaptionbox{$\gamma=1.5$}[0.47\linewidth]{\includegraphics[width=1.09\linewidth]{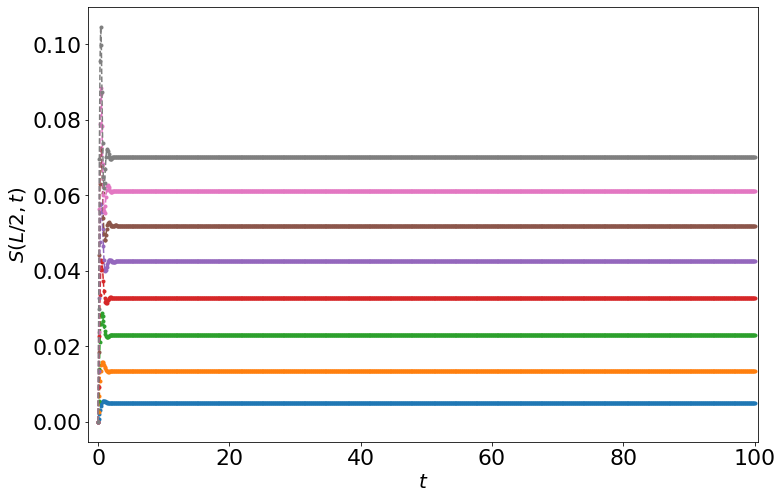}}
   \label{fig10b}
  \hfill
  \caption{Saturation value of EE versus $x$ under particle-antiparticle number filtering. } 
    \label{Fig-10.1}
\end{figure}

\textcolor{black}{From Figs.~(\ref{Fig-9}) and (\ref{Fig-10.1})} we can observe that with increasing $x$, the late-time saturation value of EE is increasing linearly with the coupling $x\,.$ This behaviour remains the same if we post-select with respect to local observables for different values of the filter rate $\gamma\,.$

\subsection{\textcolor{black}{Post-selected dynamics with respect to hopping term}}
\textcolor{black}{We now turn to the post-selected filtering dynamics associated with an extended gauge-invariant operator, namely the mesonic hopping term. This term represents the matter-gauge interaction energy of the smallest mesonic/string object, spanning two neighbouring matter sites and the intervening gauge link, and corresponds to the gauge-induced hopping term in the Hamiltonian~(\ref{H2}). Starting from the same initial state as before, we study the effective non-Hermitian dynamics generated by adding an anti-Hermitian contribution proportional to this hopping operator. This case differs from the local filtering terms considered in the previous subsection. The electric-flux operator acts on links, while the particle-antiparticle number operator acts on sites. By contrast, the mesonic hopping term acts jointly on two matter sites and the gauge link connecting them, and is therefore an extended
operator. The corresponding effective non-Hermitian Hamiltonian is
\begin{equation}
  H_{\textrm{eff}} = (x-i\gamma) \sum_{j=0}^{L-2} \left( \sigma_j^{+} \, \tau^X_{j,j+1} \, \sigma_{j+1}^{-} + \text{h.c.} \right)
+ \mu\sum_{j=0}^{L-1}(-1)^{j}\sigma^{Z}_{j}+\sum_{j=0}^{L-2}\tau^Z_{j,j+1}\,, \label{Heff43}
\end{equation}
where $\gamma$ denotes the non-Hermitian filtering strength \footnote{
\textcolor{black}{For the local electric-flux and particle-antiparticle number operators, the anti-Hermitian terms can be related, up to additive constants and rescalings, to projectors onto local eigenspaces, making the no-click interpretation relatively direct. For the non-local mesonic hopping operator $O_j=\sigma^+_j\tau^X_{j,j+1}\sigma^-_{j+1}+{\rm h.c.}$, this is not immediate: $O_j$ is Hermitian and gauge-invariant, but it is neither manifestly positive nor a projector. Hence, it is not obvious how to write the corresponding rate operator as $M_j^\dagger M_j$ for some microscopic jump operator $M_j$. This makes a direct ``measurement of the hopping operator'' interpretation ambiguous, and motivates our broader post-selected filtering interpretation. We thank the anonymous reviewer for bringing out this issue. It would be interesting in future work
to construct an explicit gauge-invariant non-local measurement channel whose no-click limit reproduces, or systematically approximates, this hopping-sector filtering
dynamics.}}.}

To compute the entanglement entropy, we bipartition the system into subsystem $A$ and its complement. Since the mesonic hopping operator acts on neighbouring sites $i$ and $i+1$ together with the intervening gauge link, it can cross the entanglement cut. In
the results presented below, we therefore mainly apply the hopping-filtering term only within subsystem $A$, excluding the term that connects $A$ to its complement from the anti-Hermitian part proportional to $i\gamma$. We then compute the EE for two representative values of the filtering strength $\gamma\,.$
\begin{figure}[htb!]
    \centering
    \subcaptionbox{$\gamma=0.4$}[0.46\linewidth]{\includegraphics[width=1.05\linewidth]{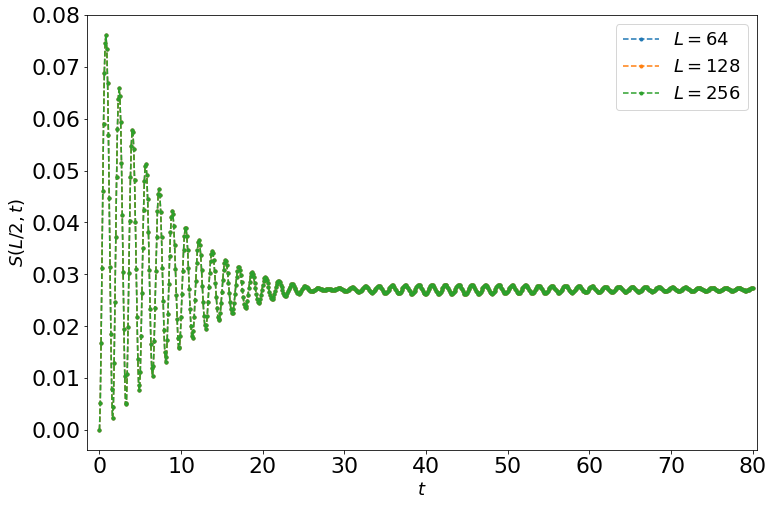}}
    \hspace{0.05\linewidth}
    \subcaptionbox{$\gamma=1.5$}[0.46\linewidth]{\includegraphics[width=1.05\linewidth]{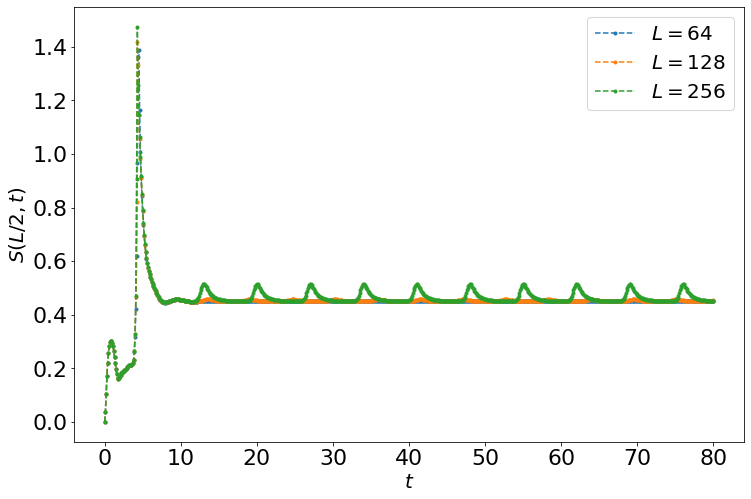}}
    \caption{Entanglement dynamics under mesonic-hopping filtering restricted to subsystem A for $x=0.5\,.$}
    \label{EE in subsystem}
\end{figure}
From Fig.~(\ref{EE in subsystem}), we observe that the saturation value of EE remains independent of the system's size even for this filtering of non-local operators. We again scan the parameter space of $\gamma$ and find that the MIPT-like transition is absent for our system, even for the \textcolor{black}{post-selection with respect to this non-local observable}  within the range of filtering strengths, evolution times, and system sizes investigated here. Also, the conclusions remain true even when we perform the post-selection by adding the hopping term throughout the system. Note that, as $\gamma$ increases, a peak in EE at early time appears (we will comment more about that shortly), although the physical interpretation of that is not clear at this moment. 

\subsection*{Dependence of late-time saturation value of EE on filtering rate:}
Having discussed the effect of non-local operators entering the filtering term on the saturation of  EE in the previous section, we make a detailed study about how this saturation value depends on the filtering strength $\gamma$ for two different coupling parameters $x\,.$ 
\begin{figure}[H]
   \centering
    \subcaptionbox{For $x=0.5\,.$}[0.47\linewidth]{\includegraphics[width=1.05\linewidth]{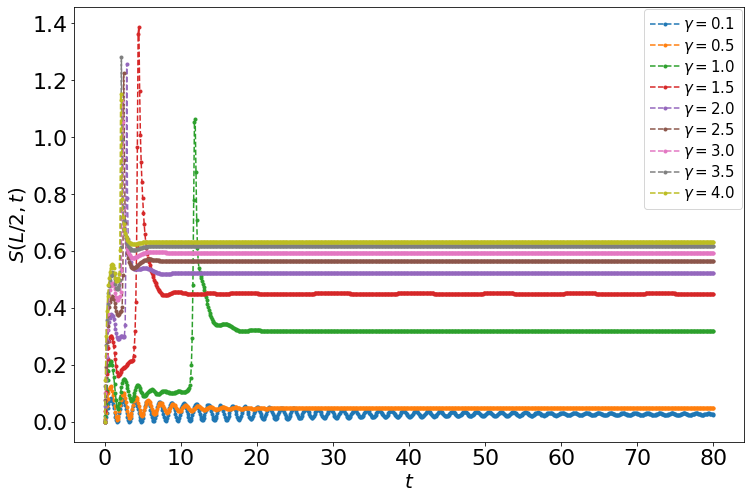}} 
   \label{fig10a}\hspace{0.10cm}
   \hfill
    \subcaptionbox{For $x=1.5\,.$}[0.47\linewidth]{\includegraphics[width=1.05\linewidth]{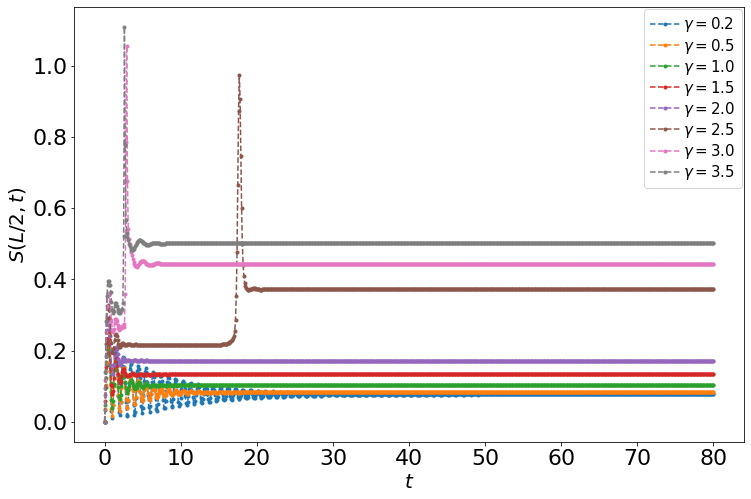}}
   \label{fig10b}
  \hfill
  \caption{Variation of the saturation value of EE with $\gamma$ under mesonic-hopping filtering.} 
    \label{Fig-10}
\end{figure}
Several interesting points can be noted from these plots in Fig.~(\ref{Fig-10}). We can summarize them below.
\begin{itemize}
\item 
  As shown in Fig.~(\ref{Fig-10}), for larger values of $\gamma$, after an initial increase, EE saturates; before that, it exhibits a peak, unlike the previous case of filtering with local operators. This indicates that the locality structure of the operator entering the non-Hermitian filtering term can significantly affect the dynamics of $S(L/2,t)$. We have checked
this feature against exact diagonalization for small system sizes and find that the peak persists, indicating that it is not a tensor-network artifact. At present, the physical origin of this peak is not fully clear, and we leave it for future study.
  \item For smaller values of $\gamma$, the EE continues to oscillate and does not saturate within the time window studied. This is expected because, in this regime, the effective Hamiltonian in Eq.~(\ref{Heff43}) remains close to the Hermitian Hamiltonian Eq.~(\ref{H2}), for which no late-time saturation of EE was observed, as discussed in Sec.~(\ref{section-3}).
  \item When $\gamma$ value is larger, EE saturates, and the late-time saturation value increases with an increase in $\gamma$. From both the plots in Fig.~(\ref{Fig-10}), we can observe that when the $\frac{x}{\gamma}$ value is less than or equal to $0.5$, peaks in EE starts to appear and the peaks start to shift towards early time ($t=0$) as $\gamma$ increases (i.e $\frac{x}{\gamma}$ decreases). 
\end{itemize}

\section{Conclusions}\label{section-6}

In this paper, motivated by recent studies of MIPT, including in post-selected no-click settings \cite{10.21468/SciPostPhys.14.5.138, 10.21468/SciPostPhysCore.6.3.051, PhysRevB.103.224210, Chakrabarti:2025hsb}, \textcolor{black}{we have initiated a study of post-selected non-Hermitian filtering dynamics in a $(1+1)$-dimensional $\mathbb{Z}_{2}$ lattice gauge theory. The filtering terms considered here are associated with gauge-invariant operators naturally present in the theory: the electric flux, the particle-antiparticle number, and the mesonic hopping operator}. At this point, we stress that our work differs from previous work, e.g., \cite{PhysRevB.111.094315}, where the authors investigate the breaking or non-breaking of Gauss's law under measurement, working in the no-click limit. In contrast, our focus is on the entanglement dynamics generated by gauge-invariant post-selected filtering terms. Since the $\mathbb{Z}_{2}$ gauge theory naturally contains both local operators and an extended mesonic hopping operator, it provides a useful setting to compare local and non-local filtering dynamics within a common framework. We also explicitly check that Gauss's law remains satisfied throughout the time evolution.

We first studied the unitary dynamics generated by the Hamiltonian \textcolor{black}{in (\ref{H2})}. In the absence of non-Hermitian filtering terms, the EE does not saturate within the time window studied. We also computed the time-averaged EE for different values of the coupling parameter $x$ and found that it increases with increasing $x$. We then turned to the \textcolor{black}{post-selected non-Hermitian dynamics}. Specifically, we investigated how the late-time behaviour of EE depends on the filtering strength $\gamma$, the system size,
and the locality structure of the operator entering the anti-Hermitian term. The numerical calculations were performed using second-order TEBD and benchmarked against exact diagonalisation, first-order TEBD, and TDVP for small systems. We also checked the time evolution using different numerical schemes, and used a second-order TEBD algorithm implemented in ITensor using MPS to control Trotter errors, especially for the extended hopping operator.

\textcolor{black}{We find that the EE saturates in the presence of the non-Hermitian filtering terms. For filtering with respect to the local electric-flux and particle-antiparticle number operators, the late-time saturation value of EE remains independent of system size for the values of $\gamma$ studied. This indicates the absence of a MIPT-like finite-size scaling transition within the range of filtering strengths, evolution times, and system sizes investigated in this paper. For the non-local mesonic hopping operator, corresponding to the smallest string-like object connecting two neighbouring matter sites through a gauge link, we similarly find that whenever the EE saturates, its late-time saturation value remains essentially independent of system size. Thus, we do not observe a MIPT-like transition (at least within the range of filtering strengths, evolution times, and system sizes considered here) in this case either.
}

We also studied the dependence of the saturation value on the filtering strength $\gamma$. For the local filtering terms, the EE initially oscillates and then saturates at late times; the oscillations are suppressed as $\gamma$ increases, showing a Quantum-Zeno-like trend. The behaviour for the non-local mesonic hopping operator is qualitatively different. Depending on the ratio $x/\gamma$, the EE can develop an early-time peak before saturating, \textcolor{black}{and the late-time saturation value increases with $\gamma$}. The appearance of this peak distinguishes the non-local filtering dynamics from the local cases. Its precise physical origin is not fully clear at present, and we leave this question for future work.

These findings open up new opportunities for future research. It will be interesting to extend this to long (mesonic) strings to see whether the EE peak appears at early times. Such a study may also clarify how string breaking is reflected in the entanglement
dynamics under post-selected filtering. One immediate thing to check is whether similar conclusions hold for $(2+1)$- dimensional $\mathbb{Z}_2$ lattice gauge theory as well. For that, the methodology developed here \cite{Bringewatt:2023xxc} might be useful. In fact, it will be interesting to perform a more fine-grained analysis after separating the total EE into a distillable and a non-distillable part \cite{Bringewatt:2023xxc}, and studying how each contribution responds to the filtering dynamics. Finally, the present work is restricted to deterministic post-selected dynamics. An
important extension would be to go beyond this limit and include the full stochastic evolution associated with monitored dynamics. It would also be useful to formulate possible circuit realizations of the filtering protocols considered here, which could
pave the way for quantum-hardware simulations. Another promising direction is to extend the present analysis to non-Abelian lattice gauge theories, such as $SU(2)$ gauge theories~\cite{Raychowdhury:2019iki}, where the richer gauge structure and larger set of physical operators may lead to qualitatively new entanglement phenomena.

\section{Acknowledgment}
Research of IR is supported by the  OPERA award (FR/SCM/11-Dec-2020/PHY) from BITS-Pilani, the Start-up Research Grant (SRG/2022/000972)  from ANRF, India, and the cross-discipline research fund (C1/23/185) from BITS-Pilani. IR acknowledges useful discussion in the meetings of QC4HEP working group. NN is supported by the CSIR fellowship
provided by the Govt. of India under the CSIR-JRF scheme (file no. 09/1031(19779)/2024-
EMR-I). NC is supported by the Director’s Fellowship of the Indian Institute of Technology
Gandhinagar. AB would like to thank the speakers and participants of the BIRS-CMI workshop (25w5386) ``Quantum Gravity and Information Theory: Modern Developments" for stimulating discussion, and the physics department of BITS-Pilani, Goa, for hospitality during this work. AB is supported by the Core Research Grant (CRG/2023/ 001120) by the Anusandhan National Research Foundation (ANRF), India.  A.B. also acknowledges the associateship program of the Indian Academy of Sciences, Bengaluru, and support from the Indian Institute of Technology Gandhinagar and a generous donor through the Singheswari and Ram Krishna Jha Chair.  NA would like to thank the Start-up Research Grant (SRG/2022/000972) for computational facilities at BITS and Sharanga HPC for tensor network computations. NA further acknowledges Sharanga HPC usage for tensor network computations, Kapil Ghadiali for computational support at TIFR, and hospitality at the Institute for Basic Science, Daejeon, where a part of the work was done. \textcolor{black}{We thank the organizers of the 42nd International Symposium on Lattice Field Theory
(LATTICE2025) for selecting the work reported in Ref.~\cite{Ara:2026o5} for an oral presentation in the parallel session on ``Quantum computing and quantum information'' \footnote{\textcolor{black}{That presentation and the corresponding proceedings focused only on the local filtering terms. The present paper substantially extends that analysis by incorporating the non-local mesonic hopping term, developing a unified post-selected filtering interpretation for both local and non-local gauge-invariant operators, and providing a more detailed study of the resulting entanglement dynamics and its scaling behavior.
Figs.~(\ref{Fig-6}), (\ref{Fig-7}), and (\ref{Fig-9}) of the present paper are refined and expanded versions of Figs.~(4), (5), and (6) of Ref.~\cite{Ara:2026o5}, respectively with additional details.}}.}
\section*{Data Availability}
The codes used for numerical simulation are available in a GitHub repository accessible from here \cite{measurement_z2_repo}\,.
\bibliographystyle{JHEP}
\appendix
\section{Numerical Convergence check}\label{app-A}
\begin{figure}[H]
    \centering
    \includegraphics[width=0.45\linewidth]{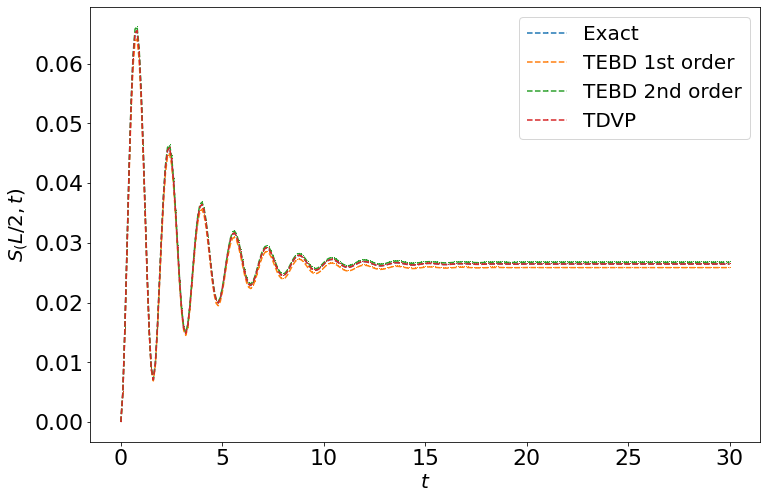}
    \caption{Entanglement dynamics for local post-selection dynamics using electric field operator for $L=8,x=0.5\,.$}
    \label{Fig-App1}
\end{figure}

To benchmark our code, we have checked our tensor network calculations with the exact diagonalisation method for a lattice site of size $L=8\,.$\\\\
\textit{Check with exact diagonalisation code:} Fig.~(\ref{Fig-App1}) shows the comparison. \textcolor{black}{We have also used three different time-evolution methods based on MPS, namely, 1st- and 2nd-order TEBD (which implement the Trotter and 2nd-order Suzuki-Trotter decompositions, respectively) \cite{itensor}, as well as the time-dependent variational principle (TDVP) \cite{PhysRevB.102.094315}\,.} As evident from Fig.~(\ref{Fig-App1}), all of them give the same results. 
In the main text, we have performed all our tensor network calculations with a cutoff ($\epsilon=10^{-8}$) and a bond dimension $\mathcal{D}=1000\,.$ To confirm the numerical convergence, we have checked our calculation for a smaller cutoff and a bigger bond dimension value. We present the results below.\\\\ 
\textit{Dependence on cut off and bond dimension:} As evident from Fig.~(\ref{bond-dimension}), our results are insensitive to these choices.
\begin{figure}[H]
    \centering
    
    \begin{subfigure}[t]{0.45\textwidth}
    \centering
\includegraphics[width=\linewidth]{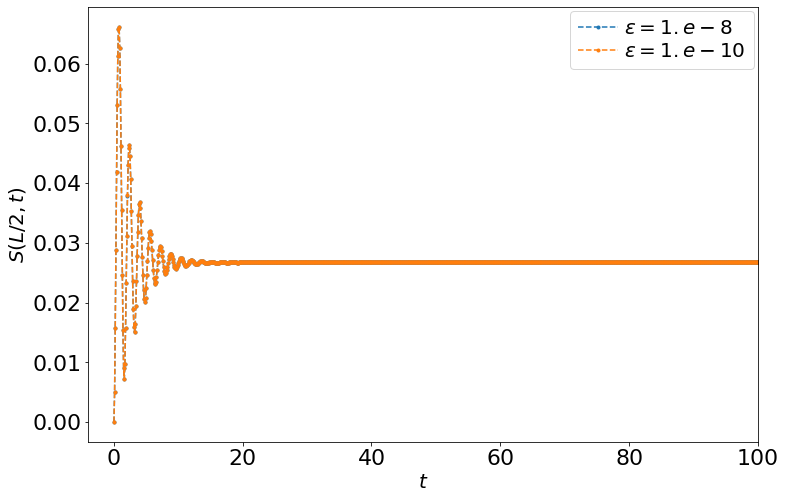}
\caption{Check for different cut-offs\,.} \label{cut-off}
\end{subfigure}
\hfill
    \begin{subfigure}[t]{0.45\textwidth}
        \centering
        \includegraphics[width=\linewidth]{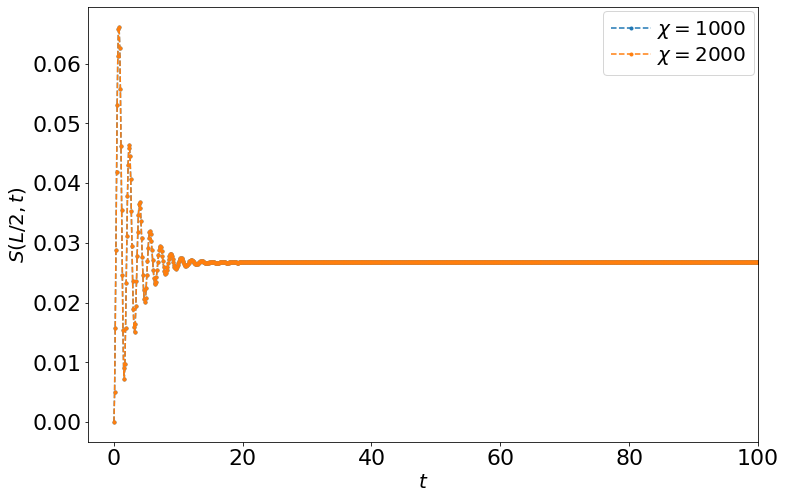}
        \caption{Check for different bond dimensions\,.} \label{EE_subs}
    \end{subfigure}

    \caption{Entanglement entropy for a system when $L=64,\gamma=0.4,x=0.5$  $(a)$  with two different cut-offs and $(b)$ with two different bond dimensions.}
    \label{bond-dimension}
  
    \label{bond-dimension}
\end{figure}
 
\section{Dependence on initial state}\label{app-B}
In the main text, we have presented results for a strongly coupled vacuum state. In this appendix, we take the initial state to be the ground state of (\ref{H2}) and show that the \textcolor{black}{non-hermitian filtering dynamics of entanglement entropy}
does not depend on the choice of the initial state. 
\begin{figure}[H]
    \centering
    \includegraphics[width=0.5\linewidth]{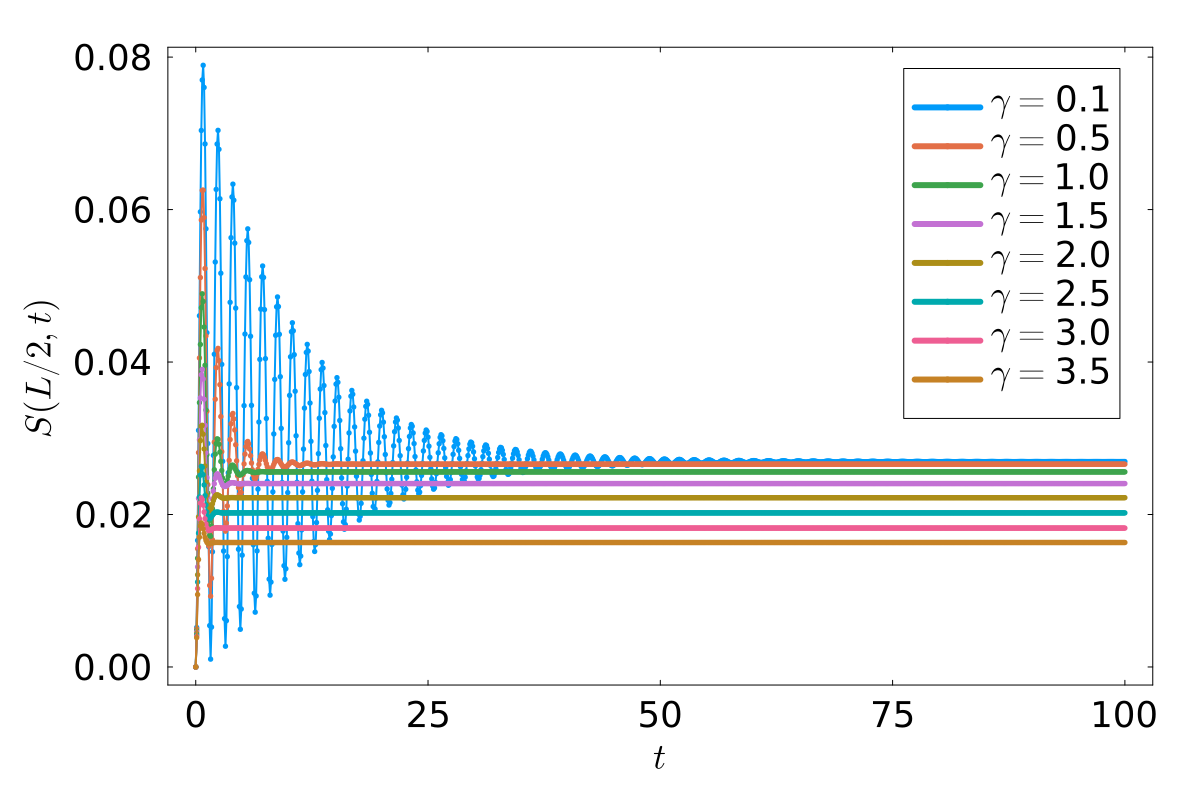}  \caption{\textcolor{black}{Entanglement dynamics under particle-antiparticle number filtering for $L=64$ and $x=1.5\,.$}}
    \label{fig:placeholder}
\end{figure}

Fig.~(\ref{fig:placeholder}) shows that with the increasing filtering rate $\gamma,$ the late-time saturation value of EE decreases. Also, we have checked the absence of MIPT-like transition.

\section{Entanglement Entropy: Dependence on $\gamma$ for $x>1$} \label{c}

 \begin{figure}[htb!]
    \centering
    \subcaptionbox{Electric flux filtering}[0.45\linewidth]{\includegraphics[width=1.09\linewidth]{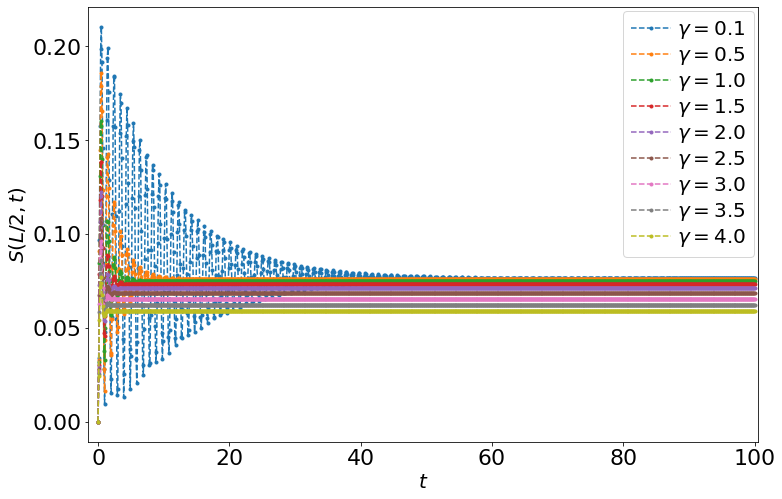}}    \label{Fig-AppD1a}
    \hspace{0.05\linewidth}\subcaptionbox{The functional form of the saturation value of EE with respect to filtering strength $\gamma\,,$ $f(\gamma)=a\gamma^{2}+b\gamma+c\,,$ with {$a=-0.000869448,\,b=-0.0010084,\, c=0.0767559$} evaluated at a late time $t_{\textrm{sat}}=100$ for all values of $\gamma$ considered here.}[0.45\linewidth]{\includegraphics[width=1.05\linewidth]{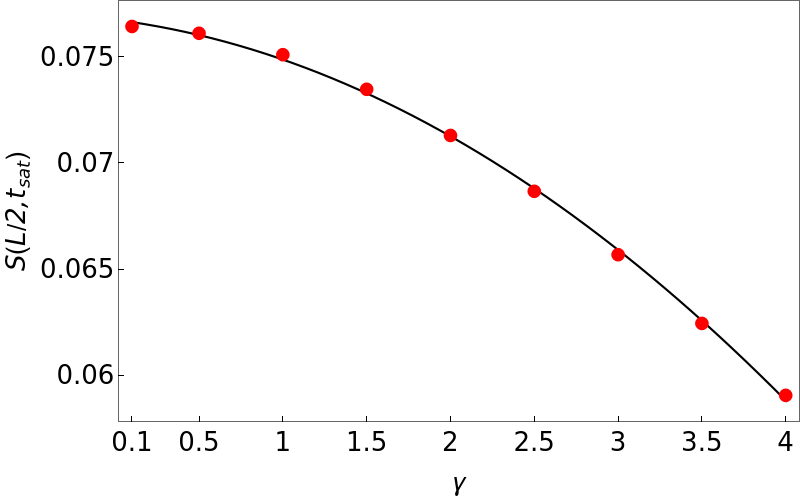}}     \hspace{0.05\linewidth}
    \caption{Entanglement dynamics under electric-flux filtering for $L=64$ and $x=1.5$.}
    \label{Fig-AppD1}
\end{figure}

In section (\ref{section-4a}), we have \textcolor{black}{studied the post-selected dynamics conditioned on the 
local physical observables}. Our computation was done for $x=0.5$ i.e $x<1$. In this appendix, we show the results for $x=1.5\,.$
 \begin{figure}[h]
    \centering
    \subcaptionbox{Particle-antiparticle number filtering}[0.45\linewidth]{\includegraphics[width=1.05\linewidth]{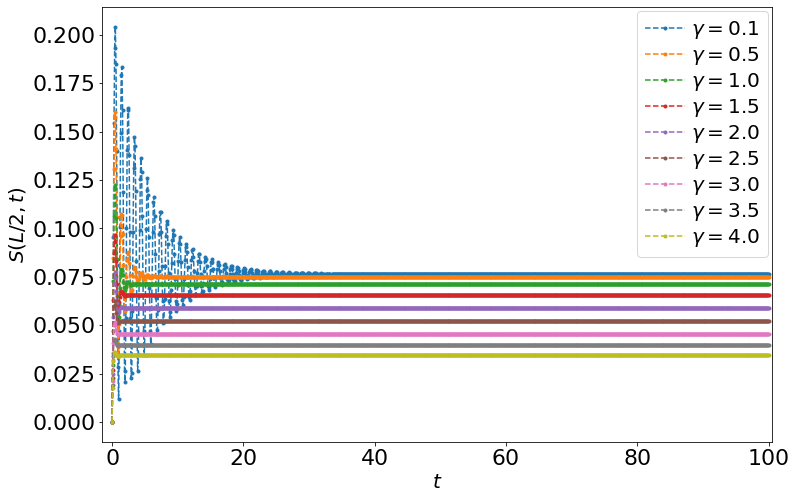}}
     \label{Fig-AppD2a}
    \hspace{0.05\linewidth} 
    \subcaptionbox{The functional form of the saturation value of EE with respect to filtering strength $\gamma\,,$ $f(\gamma)= a\gamma^{2}+ b\gamma+c\,,$ with {$a=-0.000806145,\, b=-0.00821252,\, c=0.0787908$} evaluated at a late time $t_{\textrm{sat}}=100$ for all values of $\gamma$ considered here.}[0.45\linewidth]{\includegraphics[width=1.09\linewidth]{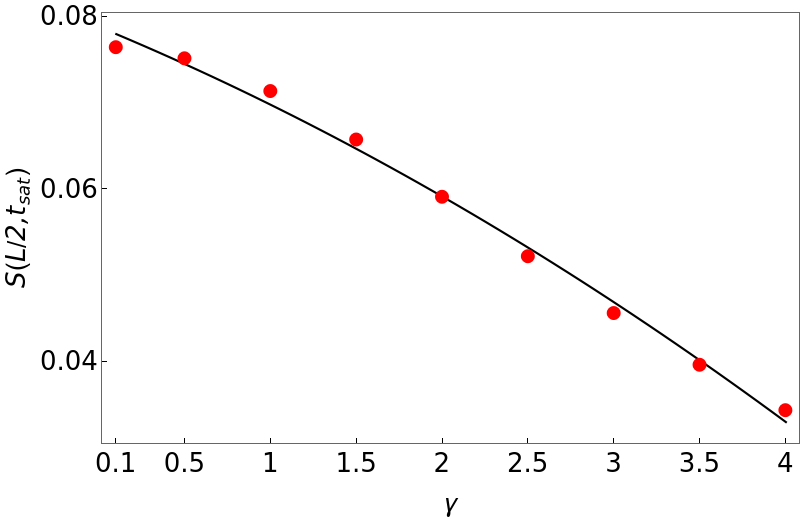}}
    \hspace{0.05\linewidth}
    \caption{\textcolor{black}{Entanglement dynamics under particle-antiparticle number filtering for $L=64$ and $x=1.5$}.}
    \label{Fig-AppD2}
\end{figure}
Again, we see that the late-time saturation value of EE decreases with respect to $\gamma$ and the behaviour shown in {\bf left panels} of Fig.~(\ref{Fig-AppD1}) and Fig.~(\ref{Fig-AppD2}) is qualitatively same as that of the corresponding Fig.~(\ref{Fig-6}) and Fig.~(\ref{Fig-7}) in the main text. However, the functional form of the late-time saturation value of $S(L/2,t_{\textrm{sat}})$ as plotted in the right panels of Figs.~(\ref{Fig-AppD1}) and (\ref{Fig-AppD2}) are different compared to their counterparts in Fig.~(\ref{Fig-6}) and Fig.~(\ref{Fig-7}) in the main text.  
\bibliography{main}

\end{document}